# Facing ADAS validation complexity with usage oriented testing

**Authors:** Laurent Raffaëlli and Frédérique Vallée, All4tec ; Guy Fayolle, Inria-Armines ; Philippe De Souza, ESI ; Xavier Rouah, Intempora ; Matthieu Pfeiffer, Magillem ; Stéphane Géronimi, PSA ; Frédéric Pétrot, TIMA ; Samia Ahiad, Valeo

## 1   Scope

Validating Advanced Driver Assistance Systems (ADAS) is a strategic issue, since such systems are becoming increasingly widespread in the automotive field. ADAS bring extra comfort to drivers, and this has become a selling point. But these functions, while useful, must not affect the general safety of the vehicle which is the manufacturer's responsibility.

A significant number of current ADAS are based on vision systems, and applications such as obstacle detection and detection of pedestrians have become essential components of functions such as automatic emergency braking. These systems that preserve and protect road users take on even more importance with the arrival of the new Euro NCAP protocols.

Therefore the robustness and reliability of ADAS functions cannot be neglected and car manufacturers need to have tools to ensure that the ADAS functions running on their vehicles operate with the utmost safety.

Furthermore, the complexity of these systems in conjunction with the nearly infinite number of parameter combinations related to the usage profile of functions based on image sensors push us to think about testing optimization methods and tool standards to support the design and validation phases of ADAS systems. The resources required for the validation using current methods make them actually less and less adapted to new active safety features, which induce very strong dependability requirements.

Today, to test the camera-based ADAS, test vehicles are equipped with these systems and are performing long hours of driving that can last for years. These tests are used to validate the use of the function and to verify its response to the requirements described in the specifications without considering the functional safety standard ISO26262.

Therefore there is also a need to improve the way of validating the ADAS functions.

## 2   The COVADEC project

The French research & development project COVADEC(*), started in the mid-2013 aims to provide methods and techniques for automotive OEMs and suppliers who face these problems.

(*) COVADEC stands for « Conception et Validation des Systèmes Embarqués d'Aide à la Conduite » which means « Design and Validation of ADAS » in French.

COVADEC main objectives are:

- Optimize test scenarios and reduce the hundreds of thousands of kilometres of driving required for the validation of ADAS functions integrated in vehicles.
- Optimize time consumption and human effort during the validation phases of ADAS.
- Meet the needs of ADAS in terms of compliance with standards such as ISO26262 or compliance with targets of occurrence of dangerous events.
- Take into account the dependability requirements upstream of the development of image processing algorithms.
- Standardize methods and tools required for the validation of functional requirements and operational safety.
- Enhance the development of ADAS functions by anticipating and implementing in priority critical situations that can default driver assistance systems.
- Ensure interoperability between test platforms, simulation platforms (PRO-SIVIC) and other development platforms (RTMaps, ADTF).

The ADAS domain is currently not properly covered by the ISO 26262 safety standard requirements, so one of the goals of the project is to propose new solutions that can fit into the validation process established for any automotive system in order to complement the ISO 26262 gaps on the subject.

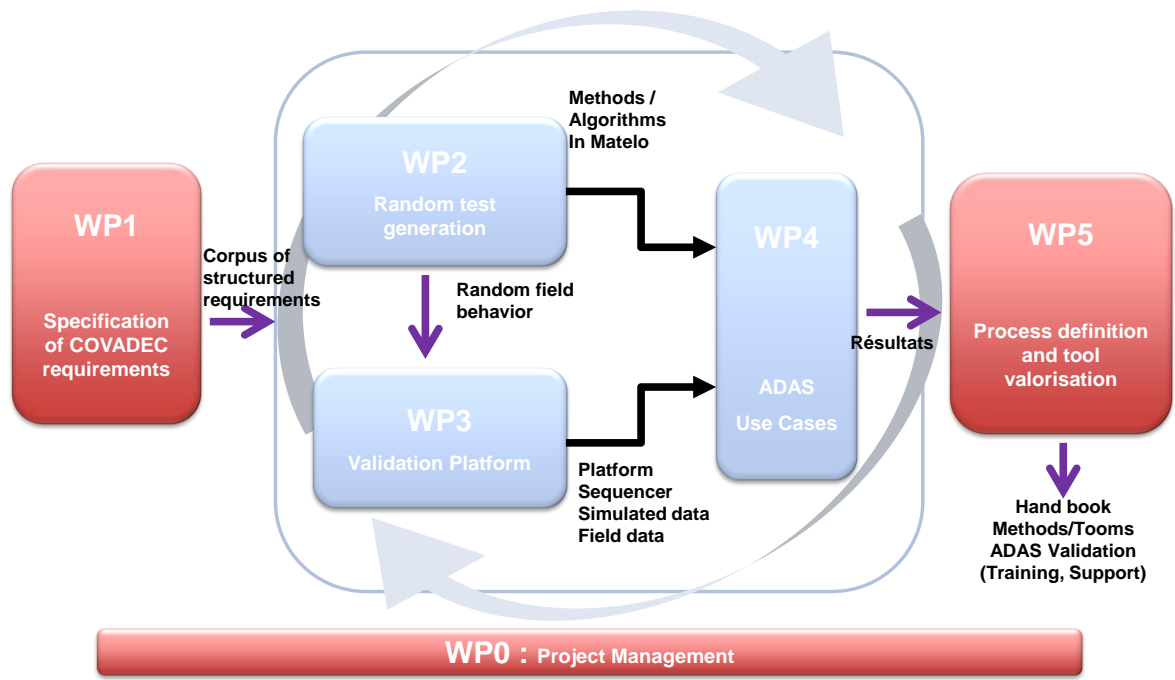

COVADEC is decomposed into 5 work packages:

- WP0: Project Management
- WP1: Specification of COVADEC requirements

The main objectives of WP1 are:

 (i) collect and synthesize the needs of stakeholders in the automotive industry in terms of certification and validation, especially for driving assistance systems based on the perception of the vehicle environment
 (ii) identify common practices of validation management and express the expectations of the industry for process improvement
 (iii) specify the architecture of the solutions proposed in response to these expectations
- WP2: Random test generation :
  WP2 aims at developing a methodology and the associated tools to generate statistical tests meeting the requirements of COVADEC project.

- WP3: Validation Platform :
  WP3 will develop and implement software tools for the execution of test cases and the generation of test reports as generated by WP2 tools.

- WP4: ADAS Use Cases :
  Two use cases have been chosen in order to demonstrate the validity of COVADEC method and platform:
  - A Lane Departure Warning (LDW) function which informs the driver from unwanted lane departure.
  - An Automatic Emergency Braking (AEB) function which prevents from front collisions.

- WP5: Process definition and tool valorisation :
  WP5 objectives are to extend the results of the case studies examined in COVADEC to all ADAS applications; to make available to the ADAS community the COVADEC results and to explain how these results will be reused and valued by COVADEC partners.

# 3 Positioning with regards to the state of the art and to ISO 26262

The stakes of functional safety, in order to guarantee safety of people and goods, is to evaluate intrinsic risks of the system and thus provide solutions to reduce the probability of hazards occurrence. The subject has been extensively covered and different standards have emerged including the ISO 26262 standard in 2011 in the automotive field, standard that can be applied to ADAS programmable electronics for automotive systems.

Regarding ADAS, the main safety risks concern an erroneous analysis of a driving situation, which may provide incorrect information to the driver or, even worse, trigger an automatic and inappropriate response of the vehicle. For instance, for the Lane Departure functions, if the ADAS is coupled to the steering control of the vehicle (Lane Keeping Assist), the system may cause the vehicle to steer unwillingly or, in the case of the Automatic Emergency Braking System, brake unwillingly.

Actually the current version of the norm ISO 26262-2011 explicitly excludes dangers that are inherent to the nominal behaviour of systems i.e. when potential hazards are caused by intrinsic limitations of sensors performances (because of their probabilistic responses), and are not a result of system failures.

After further analysis, the erroneous decisions taken by the ADAS are considered as nominal behaviour of these systems and consequently not covered by this norm. Hence, the most critical cases are not Hardware or Software failures, which are already covered by existing norms i.e. ISO 26262, but the cases in which the behaviour is diverging from the nominal behaviour. In this case, the treatment and analysis chain doesn't contain any malfunction, but produces an erroneous decision.

Such a wrongful decision implies two kinds of feared events:

- the system does not detect a dangerous situation that should have been detected as such.
- the system detects a dangerous situation when this is not the case in reality

For Automatic Emergency Braking Function, the first case is less critical in terms of functional safety, as the driver may always take the good decision and control the vehicle, while the ADAS does not trigger any action. In this case, the quality of service of the function is diminished, but there is no additional danger involved by the use of the system. On the other hand, the second case is critical, as the ADAS generates wrong information or triggers a wrong action for the vehicle. In this case, using the vehicle with the ADAS may be more dangerous than without it.

In order to carry out the validation of ADAS and their robustness against feared events, different methodologies of the state-of-the-art were investigated. Regarding verification and validation techniques of critical software, we can consider different approaches:

- The first approach, based on formal proof, has been proved to be unsuitable in the case of very complex systems (explosion of proof algorithms) or the formal expression of which is inappropriate (lower layers for example).
- The second approach based on simulation is limited by the amount of test cases that have to be generated in order to cover all possible cases.
- The third approach consists in testing of the system in real conditions. However, reaching the validation objectives in terms of testing (hundreds of millions of kilometres) is tedious or even infeasible considering the lifecycles of these systems. Furthermore, the definition and realization of test drives campaigns is proving difficult to be representative and comprehensive.

There are currently attempts to resolve these problems, for example by trying to create tool chains based on bricks such as Matlab / Simulink and / or Statemate and / or SCADE + DesignVerifier and / or Prover and / or MaTeLo and / or Teststand ... But there is today no integrated solution that addresses all of the issues raised by COVADEC (i. e. including incorporating a detailed analysis of scenarios and detection component).

The new approaches developed by COVADEC project, will give the opportunities to propose different ways to assess ADAS function and to establish safety objectives compatible with camera based systems.

In a previous paper [UCAAT 2014], we focused primarily on the COVADEC tool chain. In this paper, the main subject is to describe the process developed in order to create an efficient test database from a test model.

## 4 Methodology and technological locks

Today, there is no standard method taking into account the constraints related to the use of ADAS based on ADAS sensors such as cameras.

### 4.1 Using a statistical approach

Statistical tests as currently proposed by MaTeLo should be immersed in the ADAS context. They must be mixed with the potential of test benches and simulators environment and adapted to the analysis of the automotive dependability.

To carry out the design and validation stages for object detection systems based on cameras, numerous driving hours are required to be processed. Some cases that can be defined as critical may rarely appear or never appear during this driving campaign. We will use simulation to cover such cases. The simulation must produce synthetic image data as close as possible to reality so that the evaluated algorithms have a behaviour identical to in similar real driving conditions.

The statistical approach will also allow to address the sensitivity of the system behaviour in case of small changes of the vehicle environment. It has to be underlined, though, that many parameters defining the vehicle environment are non-independent. As a direct consequence, the values of various parameters that are exploited for the generation of tests on a statistical basis cannot be drawn completely independently. For example, the number of other vehicles (traffic density) and driver behaviours are not independent of the type of road. If the parameters are not independent, random selection may lead to generate situations that do not exist in reality, and also misrepresent the likelihood of certain situations.

The statistical approach to test generation should be supplemented with two difficulties:

- The first difficulty is a practical one: we must exhaustively know the parameters incompatibility matrix. If a driving situation can be characterized by dozens or even hundreds of parameters, knowing this matrix is not trivial.
- The second one is theoretical: the Monte Carlo method assumes that the parameters are independent, and the desired probability distribution specifically corresponds to this case. It will therefore be necessary to consider how to correct the Monte Carlo method to account for the dependence of the parameters before selecting the relevant test cases.

### 4.2 Running the test cases in the ad hoc environment

Once test cases have been identified and generated, it is necessary to be able to run them in an automated way to manage a large number of test cases. It is also necessary to reduce the time required for their implementation through the use of high-performance parallelized computing. Today there are such tools, in particular for the management of test benches, HIL (Hardware In the Loop) systems, but no one incorporating the tools dedicated to ADAS architectures.

The challenge is to provide a tool that is both easily accessible in terms of user interface and configuration (import test cases, accessibility of execution reports), modular (able to accommodate execution targets like RTMaps and Pro-SiVIC but extendable thereafter to other environments, such as Simulink) and high performance (execution distributed on multiple machines, no duplication of software resources to handle such as sensor data or 3D records that are particularly large).

### 4.3 Being able to evaluate the test results with an oracle

Another issue is to build a usable oracle, automatic and that takes into account the wide variability of situations. A secondary challenge is to determine the best location for the implementation of this oracle in the I-DEEP platform.

### 4.4 Traceability of requirements

In COVADEC the test methodology targets the verification and validation of the considered ADAS in terms of availability, reliability and functional safety. These main requirements are expressed in a restricted set of requirements, prescribing the targeted objectives of the system as rates of availability over time or reliability on detections. Hence the fulfilment of this kind of requirements can only be evaluated by taking into account the integral test campaign and the traceability of these requirements to the test sequences has no vocation to be managed in a refined manner.

On the other hand, some requirements express some behavioural rules of the ADAS when confronted to identifiable environment perturbations (e.g. inhibition rules). Then the traceability of this kind of requirements to the corresponding test sequences shall be exploited in order to provide additional information and coverage metrics for global test campaign analysis.

## 5 Test cases automatic generation

### 5.1 Problematic

ADAS validation is a complex issue, particularly for camera based systems, because these functions maybe facing a very high number of situations that can be considered as infinite. But some situations will have more influence than others on the response of the ADAS function and some will occur more frequently. So, although all situations cannot be covered by test, it is possible to reduce the space to be tested in an area that can be small enough to make test possible by choosing to test the most representative and the most influential situations that an ADAS can encounter.

Whatever the nature of data used for validation, real or simulated, the Model-Based Testing (MBT) approach can be used to automatically build a complete test database which meets these objectives of limited size while covering most of the situations that most influence the ADAS function under test.

### 5.2 Model Based Testing (MBT)

The tool used for MBT is MaTeLo (Markov Test Logic). MaTeLo is an MBT tool, which makes it possible to build a model of the expected behaviour of the system under test (SUT) and then to generate, from this model, a set of test cases suitable for particular needs (for instance, testing only the most frequently used functions of the system, or having 100% coverage of system requirements). MaTeLo is based on Markov chains. For non-deterministic generation of test cases, MaTeLo uses the Monte Carlo methods, associated with generation strategies adapted to user needs. To cope with the combinatorial explosion, we couple the graph generated by MaTeLo to an ad hoc *random scan Gibbs sampler* (RSGS), which converges at geometric speed to the target distribution as explained later in the paper. Thanks to these test acceleration techniques, MaTeLo also makes it possible to obtain a maximal coverage of system validation by using a minimum number of test cases. As a consequence, the number of driving kilometres needed to validate an ADAS is reduced.

### 5.3 Summary of the test cases generation

The following figure gives a summary of the test cases generation:

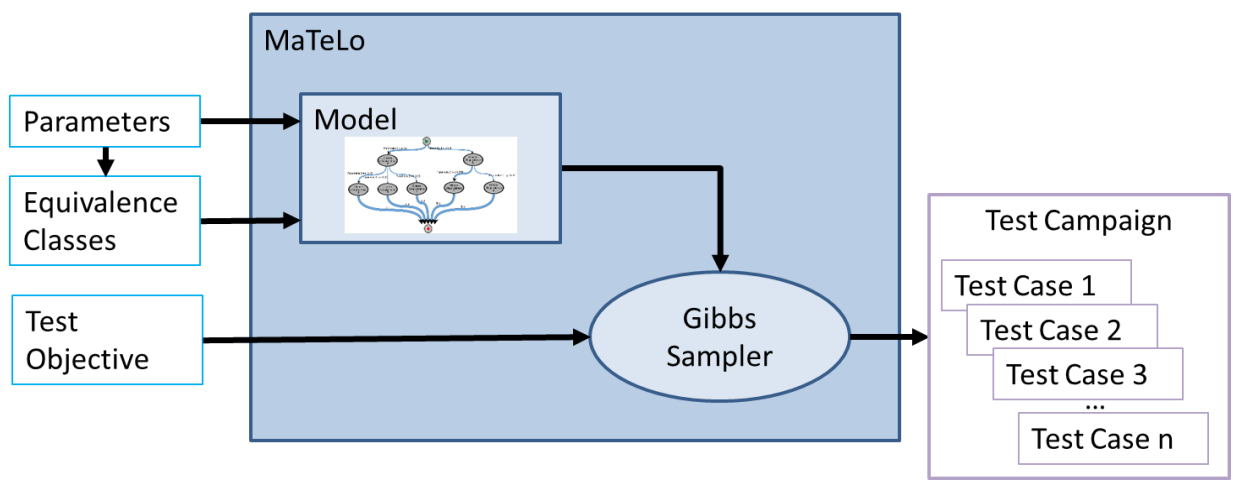

Further details are given in the following parts.

## 5.4 Global strategy

Test case generation as proposed in the MaTeLo tool faces the question of inherent combinatorial explosion. Typically, the problem is to produce samples of large random vectors, the components of which are possibly dependent and take a finite number of values with some given probabilities. One important constraint is to generate almost all situations in the most economical way. In general this task can be considered from two points of view: deterministic (via binary search trees) or stochastic, via Markov chain Monte Carlo (MCMC) sampling. In the COVADEC project, we choose the probabilistic approach, which will rely on the implementation of a Gibbs sampler, briefly described below.

- In a first step, starting from the simulation graph generated by MaTeLo, the idea is to construct a Markov random field (see section 6.4). When the parameters are locally dependent, this can be achieved by means of Bayes' formulas.
- Then test cases will be obtained by implementing Gibbs samplers. In particular, we shall strive to optimize the convergence rate toward equilibrium, since it is known from the theory that the speed of convergence is exponentially fast.

## 5.5 Gibbs samplers and random fields

In order to simulate systems with large state-space and given multi-dimensional distributions, such as those encountered in statistical physics to study equilibrium properties, powerful methods have been proposed as soon as in the 1950's. In particular, the Metropolis-Hastings's algorithms [MET], [HAST]. In the context of image processing, where digitized images can be viewed as the realization of some random field, one must quote the seminal Gibbs sampler work [GEM].

### 5.5.1 Markov Random Fields (MRF)

For an introduction to the properties of the mathematical objects presented below, the reader is referred to e.g. [GRI], [BRE].

Let $V$ denote the number of significant parameters in the system. We want to simulate the random vector $X = (X_1, X_2, \ldots X_V)$, where each component $X_i$ takes its values in a finite space $\Lambda_i$, usually called the phase space, with $|\Lambda_i| = C_i$. Typically $0 < C_i \approx 10$, and $V \approx 10^2$. The variables $X_i$ are in general dependent. Thus a configuration $x = (x_1, x_2 \ldots x_V)$ written with lowercase letters belongs to the space $\Lambda = \prod_{i=1}^{i=V} \Lambda_i$.

Of special interest will be MRF satisfying local interaction properties. This classical notion relies mainly on conditional expectation, after having defined a convenient topology on the set of indices $S = \{1, 2, \ldots V\}$ of the components of $X$, which from now on will be rather called the set of sites. Then one can define a neighbourhood system on $S$ (i.e. a topology), which is a family $F = \{N_{s \in S}\}$ such that, $\forall s \in S$,

$$s \notin N_s \text{ and } t \in N_s \Rightarrow s \in N_t.$$

The subset $\mathcal{N}_s$ is the neighbourhood of the site $s$. In a more general graph framework, $S$ is the set of vertices and $F$ defines the edges: $s$ and $t$ are linked by an edge if and only if they are neighbours, i.e. $t \in N_s$.

**Definition 1.** The random field $X$ is called a Markov random field with respect to the neighbourhood system $F$ if for all sites $s \in S$ the random variables $X_s$ and $(X_i, i \notin N_s)$, are independent given the $(X_i, i \in \mathcal{N}_s)$.

Let $\pi(.)$ denote the multivariate probability measure of the vector X, so that $\pi(x) \stackrel{\text{def}}{=} P(X = x)$. Then $\pi$ is a *Gibbs distribution* relative to the graph $\{S, F\}$ if it is of the form

$$\pi(x) = \frac{1}{Z_T} e^{-\frac{U(x)}{T}},$$

where $T > 0$ is the temperature, $U(x)$ is the energy of the configuration $x$, which derives from some potential, and $Z_T$ is the normalizing constant. Under the so-called *positivity condition* (Brook's Lemma, which is in particular satisfied when $\pi(x) > 0, \forall x \in \Lambda$), an important theorem due to Hammersley and Clifford shows the equivalence between Gibbs distributions and MRF, which in fact are essentially the same objects.

### 5.5.2 Gibbs samplers (GS)

Gibbs sampling has numerous applications and became one of the most popular routine amongst MCMC simulation methods. It applies to any multivariate distribution of the form $\pi(x_1, x_2 \ldots x_V)$. There are two main families of GS: Random scan Gibbs samplers and Periodic Gibbs samplers.

Random scan Gibbs sampler (RSGS)

The principle is simple: at each step, one selects at random a site (coordinate) $s \in S$, and then compute the new value $y_s$ of the corresponding site according to the conditional probability

$$\pi(y_s | x_j j \neq s) = \pi(y_s | x_j, j \in N_s).$$

Let $\alpha_s$ denote the probability of visiting the site s, with $0 < \alpha_s < 1$ and $\sum_1^V \alpha_s = 1$. The algorithm does construct a Markov chain $\{X(t), t = 0,1, \ldots\}$, the evolution of which is as follows.

(a) Select an initial vector X(0) and a probability vector $(\alpha_1, \alpha_2, \ldots, \alpha_V)$.
(b) On the t-th iteration,
- Choose an index *s* with probability $\alpha_s$;
- Generate $X_s(t)$ with probability $\pi(X_s | X_j (t-1), j \in N_s)$;
(c) Repeat step (b) until reaching equilibrium.

It can be shown that the Markov chain $X(t)$ is reversible, so that its invariant measure is precisely the distribution $\pi$ of the vector $X$.

Periodic Gibbs sampler

Here sites are not chosen at random, but in well-determined order fixed in advance, say $(s_1, s_2, \ldots, s_V)$ which is a permutation of $(1,2, \ldots, V)$. The algorithm generates a Markov chain Z(t) as follows. One first draws $X_{s_1}$ conditoned on the current state of the other sites, then draw $X_{s_2}$ in the same way, etc., until $X_{s_V}$. After this sweep, one says that the Markov chain Z(t) has moved exactly one step and it is not difficult to show that π is its invariant measure.

Speed of convergence

As a consequence of standard results on Markov chains, the speed of convergence to the equilibrium of the Gibbs samplers is geometric. This means that we have (see for example [BRE]) :

$$|X(0)\mathbb{P}^n - \pi| \leq \frac{1}{2}|X(0) - \pi|\delta(\mathbb{P})^n,$$

where $\mathbb{P}$ stands for the stochastic transition matrix of the Markov chain obtained from a Gibbs sampler, and $0 \leq \delta(\mathbb{P}) \leq 1$ is the Dobrushin's ergodic coefficient of $\mathbb{P}$, with

$$\delta(\mathbb{P}) = 1 - \inf_{i,j \in \Lambda} \sum_{k \in \Lambda} p_{ik} \wedge p_{kj},$$

the $p_{ik}$ 's being the elements of $\mathbb{P}$.

Computing satisfactory explicit bounds for $\delta(\mathbb{P})$ is a difficult (mostly open) problem, which depends on the kind of GS considered. For some global theoretical results in this respect, one can see for instance [LIU] and [LEV].

Indeed, the rate of convergence depends deeply on the structure of the underlying MRF describing the system. In the COVADEC project, we shall implement a RSGS. Then, by using the specific properties of the graph produced by MaTeLo, we shall analyse the speed of convergence as a function of the free probability vector $(\alpha_1, \alpha_2, \ldots, \alpha_V)$ introduced above.

## 5.6 Parameters of the MaTeLo model

In order to cover up all the situations that the ADAS systems may face, it is necessary to provide a model of the environment and driving context. The objective is to provide a meta-model of the test sequences, taking into account influential parameters that express the variability of situations the system may encounter. The construction of such a model involves taking into account parameters of heterogeneous nature, with very diverse impacts on the scene as perceived by the system.

This model must gather information about the environment in which evolves the ADAS (landscape, road type, curvature, infrastructure, etc.), driving situations (behaviour of the equipped vehicle and surrounding vehicles), weather conditions (sun, rain, fog, etc.) and known troublemakers.

The modelling of the environment needs to be as comprehensive as possible. Indeed, the model is supposed to represent any circumstances that the vehicle may encounter. Therefore, if an actually influencing parameter is forgotten, it will not be considered when creating test cases and thus the simulator may possibly never generate the situation corresponding to disturbing values for this parameter. However, these situations are potentially present in the actual video databases, even if the influential parameter in question is not explained.

We have defined several categories of influential parameters, shown below. These categories include parameters according to their nature and permit any transposition to another function.

### 5.6.1 Weather conditions

Weather conditions have an impact on how the ADAS will perceive a scene. This includes not only the weather as such, but also disturbances induced by these conditions as well as the lighting conditions of the scene.

### 5.6.2 Structure of the road and of the environment

This category includes the intrinsic characteristics of the road, that is to say the parameters to accurately describe its structure (curvature, topology, number of lanes, etc.), as well as its appearance and overall look (surface, marking, etc.).

### 5.6.3 Behaviour of the equipped vehicle

This category is used to express the behaviour of the equipped vehicle in a test sequence, both in terms of speed or trajectory rate of change. In addition, this category includes the actions of the driver that may impact the function without implying a change in the trajectory or speed (e.g. wiper operation).

### 5.6.4 Behaviour of surrounding vehicles

The presence of other vehicles can influence the perception of the scene by the ADAS either as a target vehicle or as a barrier masking what the ADAS should detect. The behaviour of other vehicles is described by a set of parameters identical to those defined for the behaviour of the equipped vehicle for which we have added parameters relating to their positioning in the scene as well as changes of trajectories they can make.

### 5.6.5 Pedestrians

This category of parameters can express how pedestrians will evolve in the scene (number, trajectory, crossing the road, etc.).

### 5.6.6 Obstacles and disturbances

This category includes all the obstacles and other disturbances known to have an impact on how the ADAS will perceive a driving situation. We grouped the barriers in several sub-categories, namely:

- Fixed Targets set on the way: this includes work pads, a stationary vehicle, a lost loading or any other object that may be on the way.
- Barriers at the trajectory limit: this includes road signs, guard rails, or a stationary vehicle.
- Pedestrians in particular situations

### 5.6.7 Equivalence classes

The range of possible values for each parameter is divided into several equivalence classes, for two main reasons:

- To select sets of values having a real impact on the ADAS function. This corresponds to the notion of "range", all situations are assumed equal within the range (e.g. 130 km / h and 131 km / h are considered equivalent in terms of ADAS, but 20 km / h belongs to a different equivalence class).
- To manage the dependencies between parameters. Indeed, some values of an influential parameter X may not be possible or have a different probability if the parameter has a value Y (Y of X correlation - examples: "night" and "sunny" are incompatible; "speed> 130 km / h" and "urban environment" is an unlikely event).

When building test campaigns, that is to say, sets of test cases which will be run for the ADAS function, if one test case has all its values in exactly the same equivalence classes another test case, it will be considered duplicate and eliminated from the campaign.

## 5.7 Structure of the MaTeLo model

The structure of the MaTeLo model to generate test cases is based on the influential parameters. In particular, the dependence between parameters is modelled as a series of dependent transitions in the MaTeLo model.

Indeed, if the parameters were independent of each other, the most natural way to build a MaTeLo model would be to create a single chain as follows:

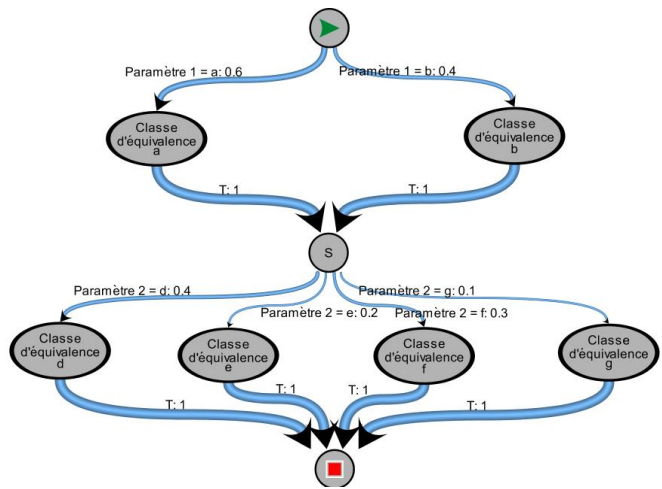

This model in the case of dependent parameters is no longer acceptable, since such model generates test cases that cannot occur in reality, distorting the representativeness of generated test campaigns.

The following graph is the graph of all possible cases of the above MaTeLo model. Let suppose that the case identified by red rectangles are impossible cases.

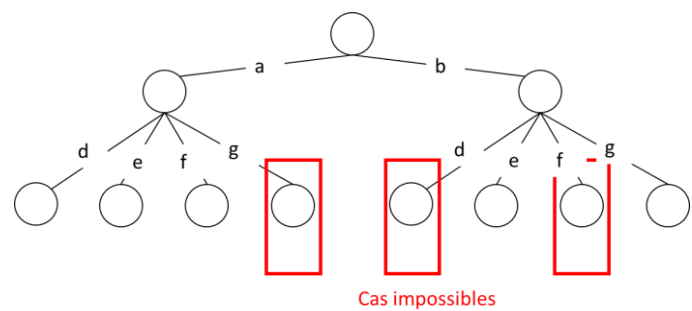

Cas impossibles

It is therefore proposed to build the MaTeLo chains as illustrated by the following example:

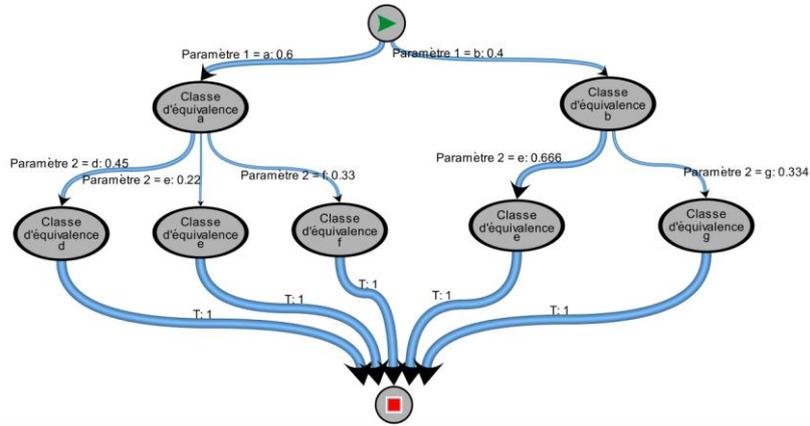

The modelled dependencies can be seen in two ways:

- Either in terms of reachable equivalence classes,
- Or in terms of probability of choosing an equivalence class knowing the equivalence class chosen for the previous parameter.

A MaTeLo model is a directed graph, so there is a notion of order to draw parameters. As a result, the dependence between parameters constrains how to build the model, including the order in which they appear.

When the level of dependencies between parameters is low (interdependencies for up to three parameters), the MaTeLo models can be simplified by creating macro-parameters, which can transform a chain of dependencies in independent chains. For instance, the moment of the day (day or night) is linked to the apparent brightness of the scene, since the brightness is darker during the night than during the day. But we could consider only the brightness, which become a macro-parameter, with a bigger range of value than considering separately either the brightness of the day or the brightness of the night. To do this, it is necessary to calculate the probabilities of the corresponding transitions, which can be obtained from the conditional probabilities that were on the initial model. Studies in the project have shown that the conditional probabilities linking the dependent parameters can be calculated from a MaTeLo model according to a simple algorithm, using Bayes formulas.

In conclusion it can be said that solutions already obtained in the project will meet satisfactory (although partially for now) to the problem of parameters dependency.

### 5.8   Parameters interpretation by the simulator

Influential parameters do not necessarily have direct translation in the simulator. A collaboration between ALL4TEC and ESI is therefore necessary to ensure that the parameters provided by MaTeLo can be properly applied in the simulator. They can be expressed in different ways in the simulator:

- thanks to existing simulation components (object models),
- through static resources called by these components (files),
- via the configuration of these components (script commands).

For performance practical aspects (related to loading times and simple definition of scenarios in the simulator), it is preferable to minimize the number of components creations and resource loads. Therefore, it is possible to create some components masking components creation or resource loading, and giving a more direct correspondence between influential parameters and script commands sufficient to describe them. The disadvantage of this approach is the need for creating a numerous components very specific to some subsets of scenarios.

## 6   Using both simulation and real data

The innovation is to manage data collection, using a statistical model based on MaTeLo tool. Instead of driving millions of kilometres aiming to encounter all function life situations, we are exploring within this project, the solution of reducing the number of kilometres by targeting the most influent conditions on the system.

A first work will be to collect sufficient real data to validate in depth the two COVADEC ADAS functions. However, in the case of some validation tests generated by MaTeLo, it is possible that the database does not cover these tests (taking into account all the various parameters / variables for this test / difficulty to test dangerous situations in real). A simulation tool must come to fill this gap in the database by synthesizing realistic sensor data for these test cases. Ideally it would be desirable that the simulation platform generates scenes, scenarios and sensor data just from the definition of the tests. In practice, certain steps must be performed in a preliminary phase before the execution of tests, in particular respective to the environment. The optimization of this process and the opportunities of generation for these items when running tests were examined in the project.

## 7 Testing tool chain for simulation

Once the test cases have been defined, it is necessary to inject them in a testing tool chain in order to execute them on the ADAS under validation. This tool chain is composed of:

- A simulator of scenarios, environments and video camera sequences (Pro-SiVIC – ESI)
- High performance ADAS data recorders (Intempora dataloggers)
- A ground-truth extraction tool for recorded data (IBEO Evaluation Suite) (VALEO)
- A framework for virtual-time or real-time execution of ADAS algorithms (RT-MAPS – Intempora)
- An ADAS hardware architecture simulator (Rabbits – TIMA)
- A test execution automation server (I-DEEP – Intempora)

Pro-SiVIC is a simulation software environment specialized in the advanced rendering of ADAS sensors (cameras, lidars, radars, GPS, communication systems…). It offers complex sensor models as well as environments taking into account numerous physical and electronic characteristics (for a camera for instance, the point is to model distortion, noise, atmospheric and climatic conditions, lighting conditions…). The key aspect for the validation process is the ability to control the characteristics of the environmental conditions. Pro-SiVIC also allows to integrate vehicle dynamic models, to setup complex driving situations in complete environments, objects animation (such as pedestrians for example). Pro-SiVIC can operate in real time or virtual time which allows addressing tests and validation use cases of ADAS functions with or without human or ECU in the loop.

RTMaps is a modular (component-based) software framework for rapid development and optimized execution of real-time applications having to manage, process and fuse numerous high-bandwidth, asynchronous and heterogeneous, sensors data streams such as cameras, lidars, radars, CAN bus, GPS, IMUs, V2V and V2I communications, etc.). RTMaps also offers data recording of any kind of ADAS sensors, then synchronized playback, in real-time or virtual time, in order to allow offline developments for perception, data fusion, communications, decision making and command-control (developments, tests, validation and benchmarking). RTMaps can also be connected to simulation and/or command control tools such as Pro-SIVIC.

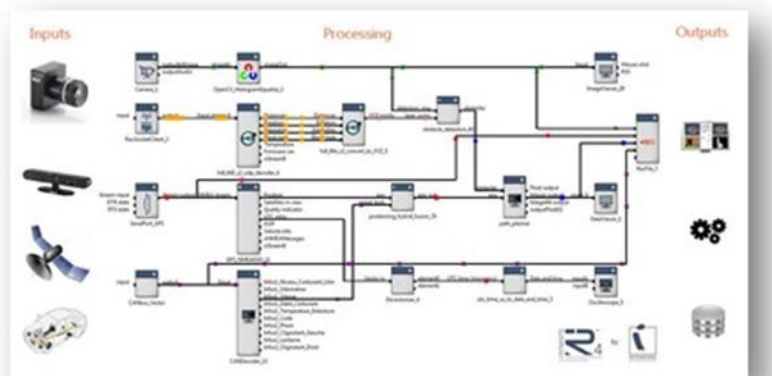

I-DEEP is a test execution automation server dedicated to validation of perception and decision making function for ADAS, particularly functions based on vision.

- I-DEEP can store recorded sensors datasets and their associated ground-truth datasets and/or simulation scenarios resources, it can as well host image processing / signal processing / data fusion algorithms to be tested (as integrated into RTMaps plugins), and then allows to define and execute automatically the numerous test cases on cluster of calculators.
- I-DEEP also offers a dual approach for validation of ADAS functions making use of simulation on the one hand and real datasets playback on the other hand. These two approaches are very complementary, simulation offering a comprehensive control of the scenario and its environmental conditions as well as the capability to test dangerous situations, whereas taking advantage of real data playback capabilities allow extension of the tests under maximum realism conditions.

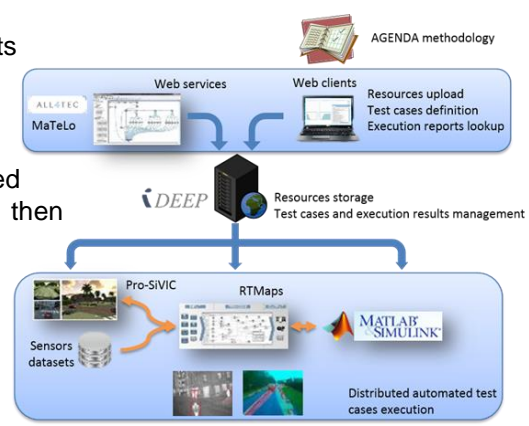

Rabbits is a fast hardware/software simulator capable of co-simulating multiprocessor systems on chip. It leverages on QEMU for processor modelling, and SystemC TLM for hardware IPs modelling. Rabbits supports many parameters, such as variable number of processors, memory size, cache availability, cache size, support of specific instructions, e.g. SIMD or floating point, etc. It also provides hardware IPs, such as memories, interrupt controller, uart, frame buffer, etc. Even though being fairly abstract, the simulation technology allows to get timing evaluation of the hardware/software system, though high level instruction execution time, instruction and data cache models, and interconnect models. In this project, Rabbits is used as a simulator of the ADAS hardware architecture with the aim of doing design space exploration of both the hardware platform and the software implementation. We did two parallel implementations of a line departure warning algorithm in C. The first one uses coarse grain (i.e. thread level) parallelism, and is executed on platforms that embed from 1 to 8 Cortex A9, leading to a factor of acceleration of 3 on the 8 core platform as compared to the unicore platform. The second implementation uses the SIMD extensions of the NEON coprocessor to express instruction level parallelism. Thanks to its capability of performing highly parallel instructions, we gained a factor of two on the already accelerated coarse grain implementation. Rabbits has been inserted in the whole design flow as an RTMaps component when targeting the validation of an optimized implementation. RTMaps pre-processes the images generated by Pro-SiVIC and sends them, though an I-DEEP interface, to Rabbits. Rabbits is concurrently running the cross-compiled LDW software that reads the images through a fake camera device hooked on the I-DEEP interface, performs the computations on them, and reports to RTMaps, through a fake serial interface also hooked on an I-DEEP interface, the status of the car on the road. RTMaps then feeds the rest of the processing chain with this information so that the appropriate decision can be taken by the system.

## 8 Expected benefits and major results

We expect benefits at many levels:

- Enhance the global knowledge of ISO 26262 applicability to design and validation of ADAS sensors, and shed light on its limitations, in order to propose solutions.
- Reduce the number of kilometres for validation of ADAS, by using a statistical model and by optimizing test plans using 'equivalence classes' principle.
- Build an ADAS validation platform (model in the Loop and software in the loop) combining real and simulation environment data.

At this stage of the project, the methodology has been entirely developed and tested on small samples of the problematic. Further developments currently in progress concern improvement of the simulator Pro-SIVIC (in order to manage a wide range of elements in the videos), of the test automation server I-DEEP (in order to use the real data from driving campaigns for tests) and the implementation of Gibbs samplers algorithms in the test case generator (MaTeLo).

Currently, a first series of tests has demonstrated a reduction in the required testing effort (considering the safety goals) by almost 90%, compared with the other available validation methods. This effort reduction target should be confirmed during the full-scale validation of the two ADAS functions expected to start from February 2016.